\documentclass[12pt]{iopart}
\usepackage{bm}
\usepackage{float}
\expandafter\let\csname equation*\endcsname\relax
\expandafter\let\csname endequation*\endcsname\relax
\usepackage{amsmath}
\usepackage{graphicx}% Include figure files
\usepackage{dcolumn}% equation table columns on decimal point
\usepackage{cite}
\usepackage{multirow}
\usepackage{footmisc}

\usepackage{soul}

\usepackage[dvipsnames]{xcolor}
\usepackage[colorlinks=true,allcolors=RoyalBlue]{hyperref}
\hypersetup{colorlinks=true}
\usepackage{amsfonts}
\usepackage[capitalize]{cleveref}
\usepackage{physics}

\newcommand{\beq}{\begin{equation}}
\newcommand{\eeq}{\end{equation}}

\newcommand{\dbar}{{\mkern+3mu\raisebox{-0.6pt}{$\mathchar'26$}\mkern-11mu {\rm d}}}

\usepackage{etoolbox}
\makeatletter
\newrobustcmd{\fixappendix}{%
  \patchcmd{\l@section}{1.5em}{7em}{}{}%
  \patchcmd{\l@subsection}{2.3em}{7em}{}{}%
}
\makeatother

\newcommand{\thetitle}{Stochastic dynamics of particles in correlated fields}

\begin{document}
\title[\thetitle]{\thetitle}
\author{Andrea Gambassi}
\address{SISSA --- International School for Advanced Studies and INFN, via Bonomea 265, 34136 Trieste, Italy}
\ead{gambassi@sissa.it}
\begin{abstract}
The effective dynamics of a colloidal particle immersed in a complex medium at equilibrium
is usually described in terms of a linear overdamped Langevin equation, possibly
with memory. However, numerical simulations and experiments have shown that this
linear model fails, suggesting that the effective dynamics of the probe is actually nonlinear.
Focusing on the case in which the medium is described by a fluctuating and
correlated Gaussian field, linearly coupled to the colloid, we derive this effective dynamics and
discuss its various consequences, including those on the stochastic thermodynamics of a
driven particle. When the field is generated by the particle itself, with negligible fluctuations,
the resulting self-chemotactic dynamics turns out to display anomalous diffusion
and run-and-tumble motion in low spatial dimension, which we characterise analytically.  
\end{abstract}

\tableofcontents
\markboth{\thetitle}{\thetitle}

\section{Introduction}
\label{sec:intro}

The stochastic dynamics of mesoscopic particles, e.g., colloids, immersed in fluctuating and possibly correlated media has long been a subject of fundamental and practical interest, as it provides a natural framework to investigate the interplay between thermal fluctuations, dissipation, and environmental correlations. 
In particular,  from the pioneering theoretical works of Einstein and Langevin~\cite{Einstein1906,Langevin_1908} and the experimental observations by Perrin \cite{Perrin_book}, it is known that, in a Newtonian fluid, the overdamped dynamics of the position $x$ of the particle is described by the linear Langevin
equation
\beq
  \overbrace{m \ddot  x}^{= 0} = -\gamma \dot x + \zeta(t),
 \label{eq:LW}
\eeq
where $\gamma$ is the instantaneous (Markovian) friction coefficient and $\zeta$ the Gaussian thermal noise, with vanishing temporal correlations. The friction coefficient $\gamma$ describes the dissipation within the medium and it is therefore expected to depend only on the properties of the fluid medium and on its coupling with the particle. In particular, one has  
either $\gamma = 6\pi\eta R$ or $\gamma = 4\pi\eta R$ depending on the boundary conditions that the particle imposes of the fluid flow (i.e., no-slip or perfect slip for a solid particle or a gas bubble, respectively), where $R$ is the radius of the particle and $\eta$ the viscosity of the medium.
In thermal equilibrium at temperature $T$, the friction coefficient $\gamma$ and the strength of the noise are related by the fluctuation-dissipation theorem \cite{Kubo1966}:
\beq
\langle \zeta(t)\zeta(t') \rangle = 2 k_{\rm B} T \gamma \delta(t-t').
\eeq
In the presence of an external potential $V(x)$, 
the Langevin equation \eqref{eq:LW} is modified by the addition on its r.h.s.~of the associated force $-V'(x(t))$, while the friction
coefficient $\gamma$ --- being solely determined by the properties of the medium and of the medium-particle interaction --- is not affected by the presence of this additional force.
This description provides a minimal, yet powerful and successful representation of thermal fluctuations and dissipation, capturing the essential features of Brownian motion in simple fluids. 

However, many realistic systems exhibit memory and non-trivial spatial and temporal correlations, leading to deviations from  standard Brownian motion. Such features arise, for instance, in complex or viscoelastic environments, active media and, more generally, or systems driven out of equilibrium, where the surrounding bath cannot be regarded as an ideal reservoir. Understanding how these correlations affect the dynamics and statistical properties of the probe particle remains a significant challenge, with implications spanning soft matter, biological systems, and non-equilibrium statistical physics. 
A notable departure from the picture outlined above arises, for instance, in the dynamics of a trapped colloidal particle probed at sufficiently high frequencies, i.e., with a high temporal resolution. This is most clearly illustrated by considering
the power spectral density $S(\omega)$ of such a motion, i.e., the Fourier transform of the stationary 
two-time correlation function $\langle x(t) x(t') \rangle$, 
as a function of the angular frequency $\omega$.  The linear equation \eqref{eq:LW} in the presence of a
harmonic trapping $V(x) = \kappa x^2/2 $ yields a Lorenzian form of $S(\omega)$, i.e., $S_0(\omega) = (k_{\rm B} T /\kappa) 2\omega_0/(\omega^2 + \omega_0^2)$, where $\omega_0 = \kappa/\gamma$ is the rate of the exponential relaxation of the overdamped particle in the harmonic trap. This prediction for $S(\omega)$ is confirmed by the experimental data (see, e.g., Ref.~\cite{Franosch2011}). 
However, if the density $\rho_f$ and the viscosity $\eta$ of the fluid medium are tuned properly, $S(\omega)$ starts deviating from the Lorentian form $S_0(\omega)$ as it develops a pronounced peak at frequencies  $\omega \simeq 1/\tau_f$ where $\tau_f = R^2\rho_f/\eta$, which is visible only if $\omega_0 \gg \tau_f^{-1}$. 
This resonance is actually due to the hydrodynamic memory, generated by the backflow of the
fluid around the moving particle \cite{Franosch2011}, which renders the effective noise $\zeta(t)$ correlated in time. This means that the simple equation \eqref{eq:LW} has to be amended by introducing a retarded friction with a memory kernel
$\Gamma(t)$, leading to
\beq
 \overbrace{m \ddot  x}^{= 0}  = -\int^t\rmd t' \, \Gamma(t-t') \dot x(t') + \zeta(t).
 \label{eq:gen-Lang}
\eeq
(While the hydrodynamic flow affects also the value of the inertial term on the l.h.s.~of this equation, the corresponding  change can be neglected for a suitable choice of time scales \cite{Franosch2011}.)
This equation is still linear but non-Markovian, due to the presence of the  memory $\Gamma$, which is still determined solely by the properties of the fluid and by the fluid-particle interaction. In three spatial dimension it is actually given by (see, e.g., Ref.~\cite{Franosch2011})
\beq
\Gamma(t)  = - \frac{\gamma}{2\sqrt{\pi}\tau_f } \left(\frac{t}{\tau_f}\right)^{-3/2} \quad \mbox{for}\quad t  \gg\tau_f.
\eeq
Since the system is at thermal equilibrium, the fluctuation-dissipation theorem implies that 
$\Gamma(t)$ determines also the correlation of the noise, i.e.,
\beq
\langle \zeta(t)\zeta(t') \rangle = k_{\rm B} T \,\Gamma(|t-t'|), 
\eeq
which also means that the noise becomes coloured.
This amended equation turns out to describe with great accuracy the experimental data for the motion of the
trapped colloidal at high frequencies \cite{Franosch2011}, which challenged the simpler model in Eq.~\eqref{eq:LW}. 
However, also the effectiveness of this generalised Langevin equation in describing the motion of tracers in fluids shows its limitations, as it was highlighted, for example, 
in Ref.~\cite{Daldrop_2017}.
In fact, the memory kernel $\Gamma(t)$, determined on the basis of the molecular dynamics simulation of a methane molecule in water and subject to harmonic confinement with strength $\kappa$, turns out to depend on $\kappa$. In particular, it was shown \cite{Daldrop_2017}  that  $\hat \gamma  = \int_0^\infty \rmd t \,\Gamma(t)$, as determined from the numerical data, monotonically increases upon increasing $\kappa$, reaching a finite value for $\kappa\to\infty$ and almost doubling compared to the plateau value which attains at small $\kappa$. 
This observation led the authors of Ref.~\cite{Daldrop_2017} to conclude 
that \emph{``the solvent friction is not determined by the solvent properties and solute-solvent interaction alone, but can be tuned by an external potential.''} 
This is clearly an undesired feature, because the basic idea behind constructing an equation of motion for the tracer particle is that the memory kernel and the friction should be determined solely by the interaction between the particle (solute) and the medium (solvent), and thus it should not depend on the possible presence of external forces (i.e., in this case, on the value of $\kappa$). 
Even more challenges
to the simple description offered by Eq.~\eqref{eq:gen-Lang} emerge when the particle is immersed in viscoelastic media. 
Consider, in fact, the experimental setting \cite{Berner_2018} in which a colloidal particle is trapped by an optical trap moving with velocity $v$ through a (viscoelastic) fluid medium. Because of the additional drag force due to $v$, the particle will displace on average from the bottom of the (harmonic) trap, climbing up the optical potential in the direction opposite to that of $v$, by an amount $\Delta x$.
What happens if the particle is further displaced (either in the direction of $v$ or opposite to it) from this position of mechanical equilibrium?\footnote{This can actually be done by post-selecting the fluctuating trajectories recorded in the stationary state on the basis of their initial position with respect to that of mechanical equilibrium \cite{Berner_2018}.}
How is it going to relax back to it?  
In the absence of driving, i.e., for $v=0$, one would expect on the basis of Eq.~\eqref{eq:LW} with the addition of the external force due to the trap, that the relaxation is a monotonic function of time $t$, i.e., an exponential with a rate set by $\omega_0$ introduced above.  This expectation is confirmed by the experiment  \cite{Berner_2018}, even if  the fluid is viscoleastic. 
However, for  $v\neq 0$, it turns out that, surprisingly, the relaxation happens via damped oscillations of the displacement  $\Delta x$, which follow the fast initial exponential decay observed also for $v=0$. 
This qualitative behaviour is actually independent of the sign of the initial displacement $\Delta x$.
It is then natural to ask whether the generalised Langevin equation \eqref{eq:gen-Lang} can account for these experimental observations. As shown in Ref.~\cite{Berner_2018}, they can indeed be reproduced, at least at a phenomenological level, by assuming that the viscoelastic medium endows the particle dynamics with a memory $\Gamma(t)$ 
consisting of the sum of two exponential contributions.
These terms have two
different time scales $\tau_1$ and $\tau_2$, with $\tau_1<\tau_2$, but also amplitudes of opposite signs,  i.e.,
\beq
\Gamma(t) = 2\gamma_\infty \delta(t) + \frac{\gamma_1}{\tau_1}\rme^{-t/\tau_1} - \frac{\gamma_2}{\tau_2}\rme^{-t/\tau_2}.
\label{eq:Gamma-fit}
\eeq
The question  at this point is whether one can understand the emergence of these various forms of effective dynamics (as well as memory kernels) for the tracer particle within a unifying picture or via simplified models. 

Keeping in mind that the evolution equation for a particle in a medium actually results from coupling the particle to the medium and from tracing the latter out, the first step is to model the medium. In order to reduce the complication of the task, this is usually done by replacing it with a number of fictitious ``bath'' particles of masses $m_i$, coordinates and momenta $(q_i,p_i)$, with $i=1, \ldots, N$,  which are assumed to interact with the tracer particle via a potential $V_{\rm int}(x-q_i)$. The tracer has mass $m$, coordinate $x$, momentum $p$, and is subject to an external potential $V_{\rm ext}(x)$. The parameters and number $N$ of these fictitious particles are then fixed on the basis of physical intuition and the expected behavior of the medium. 
Once this is done, this problem of classical Hamiltonian dynamics cannot be generically solved but it can be dealt with (also in the case of overdamped dynamics) by using, for example, the Mori-Zwanzig
projection operator formalism \cite{Mori_1965, Zwanzig_book}. 
The latter allows one to derive an effective, generally non-Markovian, but linear Langevin equation, characterised by a certain memory kernel $\Gamma(t)$, which can be calculated. 
If $V_{\rm int}(x)$ is quadratic, the problem can be solved exactly.  
In some cases, even the model involving only one bath particle, i.e., with $N=1$, is sufficient to
reproduce some of the observations discussed above. 
For example, the dependence of the integrated memory $\hat \gamma$ on the strength $\kappa$ of the optical trap which was reported  in Ref.~\cite{Daldrop_2017}, can be qualitatively reproduced \cite{Muller_2020} 
by assuming $V_{\rm int}(x) = V_0\cos (2\pi x/d_0)$ (the so-called Prandtl-Tomlinson model for stick-slip motion).
The approach briefly outlined above, however, does not consider the possible spatial structure of the medium, which is completely neglected when it is replaced by a collection of fictitious particles. This means, inter alia, that one cannot explore aspects such as the spatially-resolved energy flows within the medium or the possible fluctuation-induced forces between particles which are simultaneously present within the same bath. Similarly, the effect of extended spatial correlations of fluctuations across the medium are out of reach of that approach. 
Example of media in which these
aspects cannot certainly be neglected but are actually of interest are provided by near-critical liquid media: upon
approaching their critical point, a collective fluctuating field naturally emerges, which is
the order parameter $\phi$ of the phase transition. The order parameter develops correlated fluctuations both in space and in time, and the spatial
correlation length $\xi$ and corresponding correlation time might become so large as to
challenge the separation of time scales, which is usually behind the idea of deriving a
Langevin equation. 
Moreover,  in these critical media, it was theoretically predicted and experimentally demonstrated 
that when the correlation length $\xi$ becomes comparable with their distance, two particles are subject to a 
mutual fluctuation-induced force, known as critical Casimir forces (see, e.g., Refs.~\cite{Gambassi_2009,gambassi_critical_2024} for reviews) which were also mentioned in the motivation for the award to Mehran Kardar of the 2025 Boltzmann Medal. 
In the remaining part of this work, we briefly review some recent progresses in understanding the stochastic dynamics of particles in correlated fields, which we contributed to. In particular, in Sec.~\ref{sec:model} we introduce a simple model of a particle in interaction with a fluctuating Gaussian field $\phi$, both of which are in contact with a bath at thermal equilibrium. We present the effective dynamics of this particle and the fluctuation-dissipation theorem that it satisfies. Then, in Sec.~\ref{sec:eq-dyn} we discuss the features of this effective dynamics, focusing on the power spectral density of the fluctuations and of their non-Gaussian statistics. In Sec.~\ref{sec:driving}, the particle is driven out of equilibrium by a moving trap. Similarly to the case of viscoelastic media discussed above, it turns out that the relaxation of the particle after a displacement occurs via damped oscillations. In this non-equilibrium setting and due to the presence of spatial correlations, the stochastic thermodynamics of the system needs to be formulated in terms of the fields associated with work and heat. Depending on the correlation length $\xi$, for example, it turns out that the rate at which the field $\phi$ extracts heat from the thermal bath acquires an unexpected spatial structure. In Sec.~\ref{sec:active} we consider a non-reciprocal extension of the model introduced in Sec.~\ref{sec:model} which actually describes the behaviour of self-chemotactic particles. In particular, we focus on how the mean  square displacement at long times depends on the dimensionality of the problem and on the possible attractive or repulsive character of the self chemotaxis. Finally, in Sec.~\ref{sec:outlook} we present our summary and outlook. 

\section{A simple model}
\label{sec:model}

\begin{figure}
    \centering
  \includegraphics[width=0.85\linewidth]{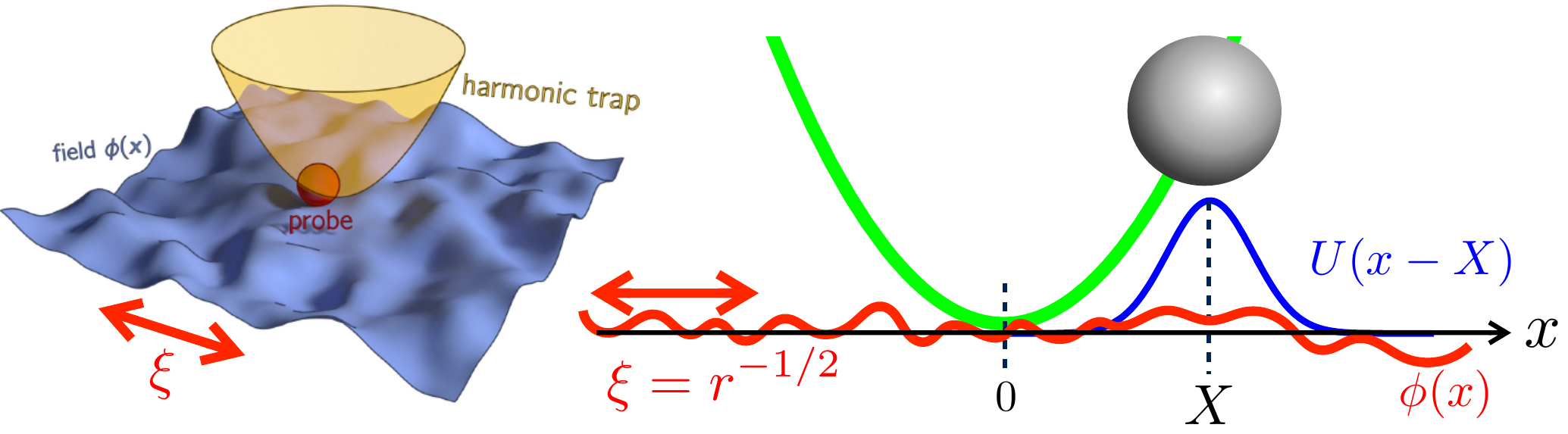}
% \put(-315,-3){(a)}
%\put(-125,-3){(b)}
\caption{Left: Schematic representation of a colloidal probe confined by an optical harmonic trap and interacting with a  fluctuating field $\phi$, the fluctuations of which are correlated across a distance set by the correlation length $\xi$.  Right: The colloidal probe at a certain position $X$ is modelled by a particle (gray) in linear interaction with the field $\phi(x)$ (red) at position $x$ via a potential $U(x-X)$ (blue), see Eq.~\eqref{eq:model}. 
The particle is also subject to a harmonic potential (green), with the minimum at $X=0$. %
(The left panel is adapted from Ref.~\cite{Basu_2022}.)}
\label{fig:cartoon-a}
\end{figure}

Inspired and motivated by the state of affairs described in Sec.~\ref{sec:intro} and by previous works \cite{Demery_2010,Demery_2010_2,Dean_2011,Demery_2011,Demery_2013}, 
we considered  a simple model \cite{Basu_2022} in
which the probe particle is confined by a harmonic potential while being in interaction 
with a thermally fluctuating Gaussian field $\phi$, the fluctuations of which are characterised by a certain correlation length $\xi$, see the cartoon on the left of Fig.~\ref{fig:cartoon-a}.
The model is described by the effective Hamiltonian
\beq
{\cal H}[\phi,X] = \int\!\rmd^d x \left[ \frac{1}{2} \left( \nabla\phi\right)^2 + \frac{r}{2}\phi^2\right] + \frac{\kappa}{2} X^2  - \lambda \int \!\rmd^d x  \, \phi(x) U(X-x),
\label{eq:model}
\eeq
where $d$ is the spatial dimensionality of the system.
Accordingly, the fluctuating field $\phi$, in equilibrium, has a correlation length $\xi$, which is related to the parameter $r>0$ as 
$\xi = r^{-1/2}$, where $r=0$ corresponds to the critical point for the field. The particle, with coordinate $X$, is subject to 
an optical potential $\kappa X^2/2$ and is coupled to the fluctuating field via the last term in Eq.~\eqref{eq:model}, which is linear in $\phi$ (but \emph{not} in $X$) and involves the potential $\lambda U$. The coupling constant $\lambda$ is introduced here as a bookkeeping for the perturbative expansions discussed further below. A more detailed cartoon of the model is presented in the right of Fig.~\ref{fig:cartoon-a}. 
For example, if $U$ is a delta function, then the linear interaction term above is just equal to $-\lambda \phi(X)$, i.e., to the value of the field at the position of the particle. 
Note that, a major simplification compared to the cases discussed in Sec.~\ref{sec:intro} is that the fluctuating medium represented by $\phi$ can occupy the volume of space corresponding to the particle, which is actually solely represented by its coordinate $X$ and by the interaction potential $U$. 
The stochastic dynamics of the particle and the field can now be specified on the basis of ${\cal H}[\phi,X]$ in Eq.~\eqref{eq:model}: 
for example, assuming a dynamics for the field $\phi$ which satisfies a local conservation law (the so-called model B \cite{Halperin_1977}), one naturally writes:
\beq
\begin{cases}
{\displaystyle \frac{\partial}{\partial t}} \phi(x,t) = D \nabla^2 {\displaystyle \frac{\delta {\cal H}[\phi,X]}{\delta\phi(x,t)}} - \nabla\cdot\eta(x,t),\\[3mm]
\underbrace{m \ddot  X}_{= 0}  = -\gamma_\infty \dot{X}  - \nabla_X {\cal H}[\phi,X] + \zeta(t).
\end{cases}
\label{eq:dynamics-p-f}
\eeq
The dynamics of the particle is assumed to be overdamped where, in addition to the friction force, one accounts also for the force $ - \nabla_X {\cal H}[\phi,X] $ deriving from the interaction of the particle with the field in Eq.~\eqref{eq:model}.
Note that $\gamma_\infty$ in the previous equation is the friction coefficient that the particle would experience in the medium in the absence of the coupling to the field $\phi$, i.e., for $U=0$. In this respect, we emphasise that $\phi$ has to be understood as an additional and emerging collective mode of the underlying medium, as in the case in which $\phi$ represents the order parameter of a near-critical liquid. %
%
%%
%%
%%
%%
%%%%%%%%%%%%%%%%%%%%%%%%%%%%%%%%%%%%%%%%%%
%%%%%%%%%%%%%%%%%%%%%%%%%%%%%%%%%%%
%%
%%
%%
The thermal noises $\eta(x,t)$ and $\zeta(t)$ in Eq.~\eqref{eq:dynamics-p-f} 
are assumed to satisfy the fluctuation-dissipation relations
\beq
\begin{cases}
\langle \eta_i(x,t) \eta_j(x',t')\rangle = 2 D T \delta_{ij}\delta^d(x-x')\delta(t-t'),\\
\langle \zeta_i(t) \zeta_j(t')\rangle = 2 \gamma_\infty T \delta_{ij}\delta(t-t'),
\end{cases}
\label{eq:noises}
\eeq
because both the particle and the field are in equilibrium with the environment at temperature $T$. 
Due to the choice of a Gaussian field and of a linear dependence of the particle-field coupling on $\phi$, the equation of motion for $\phi$ is linear and thus it can be solved. Its solution can then be used in the equation of motion for the particle. As a result, the last two terms on the r.h.s.~of the second equation in Eq.~\eqref{eq:dynamics-p-f} changes according to \cite{Basu_2022}
\beq
- \nabla_X {\cal H}[\phi,X] + \zeta(t) \mapsto  -\kappa X + \int_{-\infty}^t\!\!\rmd t'\, F(t-t', X(t)-X(t')) + \Xi(x,t),
\label{eq:eff-force}
\eeq
where
\beq
F_l(t,x) =\lambda^2D \int\frac{\rmd^dq}{(2\pi)^d} i q_l q^2 |U_q|^2\rme^{i q\cdot x - Dq^2(q^2+r)t}, 
\label{eq:eff-force-exp}
\eeq
and 
\beq
\langle \Xi_l(x,t)\Xi_m(x',t')\rangle = T  \left[ 2\gamma_\infty\delta(t-t')\delta^d(x-x')\delta_{l,m} + G_{lm}(x-x',t-t')\right].
\label{eq:corr-noise}
\eeq
Note that the force on the r.h.s.~of Eq.~\eqref{eq:eff-force} is not a linear function of $X(t)$, because $F$ in Eq.~\eqref{eq:eff-force-exp} is non-linear, in contrast with the forces in the Langevin equations considered in Sec.~\ref{sec:intro}. Moreover, this force depends on the difference $X(t)-X(t') = \int_{t'}^t\rmd t'' \dot{X}(t'')$ and thus the dynamics is not only non-linear but also generically non-Markovian. Due to the fact that $F(t,x)$ vanishes for $x=0$ (see Eq.~\eqref{eq:eff-force-exp}), the friction force depends linearly on $\dot X(t')$ for small velocities $\dot X(t')$  and thus it can be interpreted as a generalised non-linear friction characterised by a memory kernel.

The noise $\Xi(x,t)$ in Eq.~\eqref{eq:eff-force} turns out to be coloured with space-dependent correlations, as shown by Eq.~\eqref{eq:corr-noise}.  Nonetheless, it is connected to the dissipation represented by the non-linear memory kernel $F(t,x)$ 
 by a generalized form of the fluctuation-dissipation theorem \cite{Basu_2022}, which takes the form
 \beq
\frac{\partial F_m(t,x)}{\partial x_l} = - \partial_t G_{lm}(t,x) \quad \mbox{for}\quad t>0,
\eeq
with $G(x,-t) = G(x,t)$.

\section{Dynamics at equilibrium}
\label{sec:eq-dyn}

\begin{figure}
    \centering
  \includegraphics[width=0.85\linewidth]{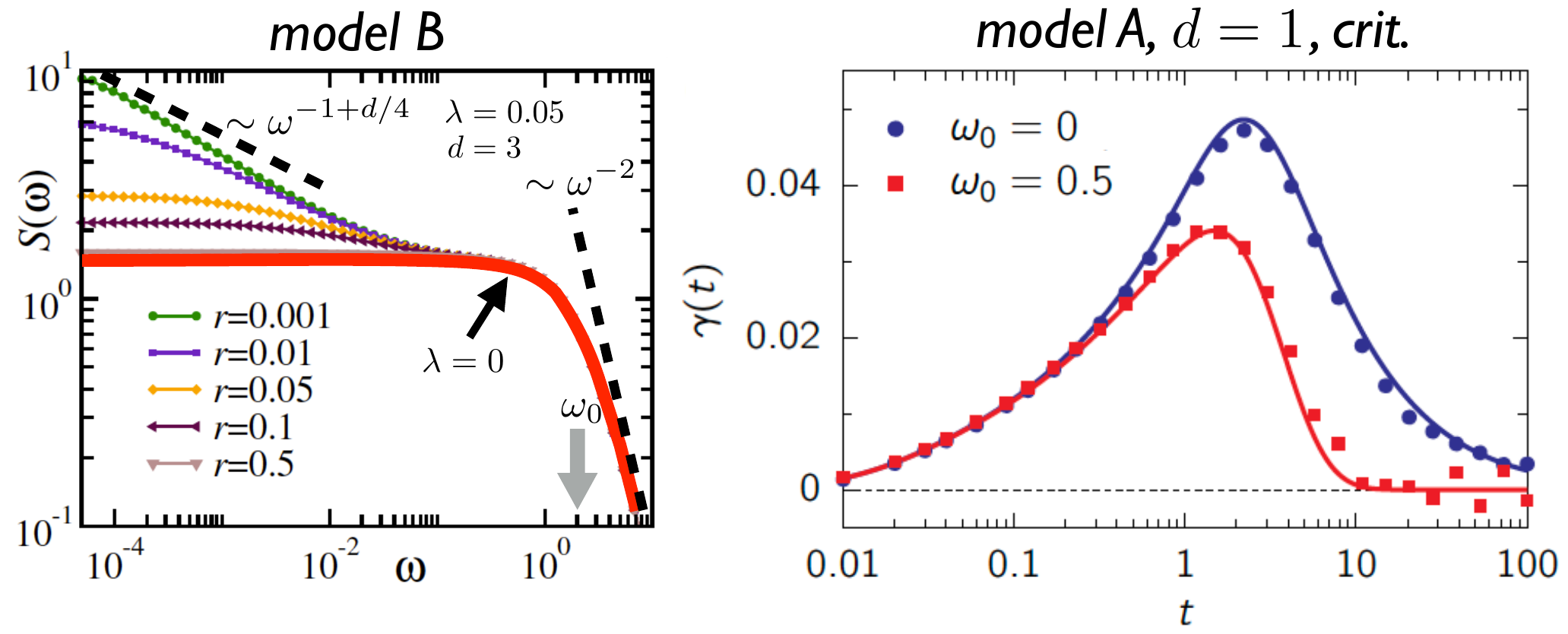}
%   \put(-390,4){(a)}
%  \put(-185,4){(b)}
\caption{Left: Power spectral density $S(\omega)$ of the position of the particle in a field with model B dynamics in $d=3$ and various values of the distance $r$ from the critical point $r=0$, as obtained analytically from a perturbative calculation at order $\lambda^2$ (with $\lambda=0.05$ for illustration). Upon approaching the critical point, the algebraic behaviour $\sim \omega^{-1+d/4}$ emerges for $\omega\to 0$ (dashed line on the left) , while the behaviour $\sim \omega^{-2}$ for $\omega \gg \omega_0$ (dashed line on the right) is not affected by the coupling to the field. Here a Gaussian interaction potential $U(x) = \exp(-x^2/(2R))$ is used with $R=2$, while the other parameters except $\kappa=2$ are set to 1. (This panel is adapted from Ref.~\cite{Basu_2022}).  
Right: Excess kurtosis $\gamma(t)$ (see Eq.~\eqref{eq:exc-k}) of the displacement $Y(t)$ as a function of time $t$ for a critical Gaussian field with model A dynamics, in spatial dimension $d=1$, in the presence (red line and symbols) or in the absence (blue line and symbol) of a harmonic trap with relaxation rate $\omega_0$ (see the main text). The solid lines correspond to the analytical prediction based on perturbation theory at order $\lambda^2$ (with $\lambda = 0.5$ for illustration), in good agreement  
with the results of numerical simulations of the dynamics (symbols), beyond perturbation theory. In this panel the potential  $U(x) = \delta(x)$ was assumed, while all other relevant parameters were set to 1. (This panel is adapted from Ref.~\cite{Demery_Gabassi_2023}.)}
\label{fig:dyn-eq}
\end{figure}

Having established the effective equation of motion for the particle in interaction with the field, one can now explore the observable consequences of having both a \emph{non-linear} and \emph{non-Markovian} evolution.  As we discussed in Sec.~\ref{sec:intro}, the power spectral density $S(\omega)$ (i.e., the Fourier transform of the two-time correlation function $\langle X(t)X(t')\rangle$ in the stationary state) provides useful information about the dynamics. 
In the absence of the interaction between the particle and the field, i.e., for $\lambda=0$, this is given by the Lorentzian form discussed in Sec.~\ref{sec:intro} (with $\omega_0=\kappa/\gamma_\infty$) and shown as a red line in the left panel of Fig.~\ref{fig:dyn-eq}.
Upon setting $\lambda \neq 0$, the evolution equation, being non-linear, can no longer be solved exactly. A viable approach (which can be tested against numerical simulations) is to resort to a perturbative expansion in $\lambda$, which turns out to involve only even power of the coupling, due to the symmetry $\{\phi,\lambda\} \leftrightarrow \{-\phi,-\lambda\} $  of the problem \cite{Venturelli_2022,Basu_2022}. 
The resulting spectral density $S(\omega)$, including the first non-trivial correction at order $\lambda^2$, is also shown in the left panel of Fig.~\ref{fig:dyn-eq} in spatial dimension $d=3$ and for various values of the distance $r$ from the critical point (located at $r=0$).
One observes that, upon approaching the critical point, $S(\omega)$ develops an algebraic singularity $S(\omega) \propto \omega^{-1+d/4}$ as $\omega \to 0$ (we assume $d<4$). This behaviour differs qualitatively from the one due to the hydrodynamic memory discussed in Sec.~\ref{sec:intro}, which takes the form of a resonance occurring at a frequency determined by the properties of the medium. Here, instead, the modification occurs at low frequencies, as heuristically expected whenever a critical behaviour is involved. In particular, this implies that the effective noise becomes coloured due to the emerging memory. 
%

%%
%%
%%
%%
%%%%%%%%%%%%%%%%%%%%%%%%%%%%%%%%%%%%%%%%%%
%%%%%%%%%%%%%%%%%%%%%%%%%%%%%%%%%%%
%%
%%
%%

Another quantity which provides information on the possible non-Markovian and non-linear nature of the dynamics and which is experimentally accessible is the statistics of the 
fluctuations of the displacement $Y(t) = X(t)-X(0)$ in the equilibrium stationary state. 
Note that both $X(t)$ and $X(0)$ are separately characterised by Gaussian fluctuations in equilibrium: the marginal of the Boltzmann distribution $\propto \exp\{-{\cal H}[\phi,X]\}$ with ${\cal H}[\phi,X]$ given in Eq.~\eqref{eq:model} turns out to be  $\propto \exp\{-\kappa X^2/(2T)\}$ \cite{Basu_2022}. Accordingly, if the time $t$ is longer than the longest relaxation time within the system, these two variables are uncorrelated and $Y(t)$ is Gaussian. However, this is not the case at smaller $t$, and deviations from the Gaussian distribution  can be detected  \cite{Demery_Gabassi_2023} by focussing on the excess kurtosis  
\beq
\gamma(t) = \frac{\langle Y^4(t)\rangle}{\langle Y^2(t)\rangle^2}-3,
\label{eq:exc-k}
\eeq
which vanishes for a Gaussian distribution. 
As for the structure factor, $\gamma(t)$ can be calculated in a perturbative expansion. The right panel of Fig.~\ref{fig:dyn-eq} shows the evolution of $\gamma(t)$ at criticality and in $d=1$, assuming that the dynamics of the field does not obey the local conservation law, i.e., that it is given by the so-called model A \cite{Halperin_1977} instead of the model B in Eq.~\eqref{eq:dynamics-p-f}. The two curves correspond to different values of the relaxation rate $\omega_0$ in the trap, with $\omega_0=0$ representing diffusion in the absence of confinement.  As expected, $\gamma(t)$ vanishes at long times,  but it features a prominent peak at a certain finite time $t=t^*$, at which the deviation from the Gaussian statistics  is maximal. Correspondingly, one can show \cite{Demery_Gabassi_2023} that the probability density  of $Y(t^*)$ is well approximated by a Gaussian for small values of $Y(t^*)$, while it departs from it in the tails of the distribution.
In passing, we mention that similar deviations have been observed experimentally in viscoelastic media \cite{Muller_2020}. 
These departures from Gaussianity indicate that the effective equation of motion for the particle is  
actually non-linear. 
Additional effects of the coupling to the field can be seen by studying the relaxation of the particle towards the position $X=0$ of mechanical equilibrium in the trap after a displacement away from it, with $X(t=0)\neq 0$. In this case, one finds \cite{Venturelli_2022}  that $X(t)$ shows a fast exponential decay with a rate set essentially by $\omega_0$ 
due to the presence of the trap, 
followed by a long-time algebraic decay. This eventual algebraic law is due to the fact that
the field slows down the particle, either because the field relaxes slowly close to the critical point (the so-called critical slowing down) or because it obeys generically a local conservation law (model B).

\section{Driving the particle}
\label{sec:driving}

In the previous section, we discussed the case in which the particle and the field are both in contact with an equilibrium thermal bath, focussing on various aspects of the resulting effective dynamics of the particle. However, the simple model introduced in Sec.~\ref{sec:model} can be easily generalised to encompass the relevant case in which the particle is driven through the medium by a moving trap, e.g., a moving optical tweezer (such as that experimentally realised in, e.g., Ref.~\cite{Berner_2018}). 
Differently from the previous case, the system constituted by the particle and the field will eventually settle in a non-equilibrium steady state, displaying the novel behaviours discussed below, some of which resemble those mentioned in Sec.~\ref{sec:intro}, though in a different physical setting.

\subsection{Oscillations in an overdamped dynamics}

The presence of a trap moving with velocity $v$ and driving the particle 
can be accounted for  \cite{Venturelli_2023} by 
replacing 
\beq
X \mapsto X - vt,
\label{eq:map-v}
\eeq
in the expression of the potential due to the optical trap, i.e.,
in the second term on the r.h.s.~of 
Eq.~\eqref{eq:model}.
This means that the minimum of the harmonic trap moves as $vt$ in space. As a consequence of the additional drive experienced by the particle, its mean position $\langle X\rangle$ grows linearly as $\langle X\rangle = v t + X_\infty$, where the offset $X_\infty$ is determined by the balance between the friction due to the velocity $v$ and the restoring force exerted by the moving trap. In particular, in the absence of the interaction with the field $\lambda=0$, this balance yields $X_\infty = - v/\omega_0$, which is corrected in the presence of the field. Note that $X_\infty$ actually represents the position of mechanical equilibrium of the particle in the frame of reference co-moving with the trap.  Moreover, while being driven, the particle affects the field $\phi$ around it in a way that is no longer isotropic in average (as it was in equilibrium)  but reflects the fact that the particle is driven in a certain direction. 
For convenience, in the discussion below, we refer the position $X$ to the co-moving frame centred at $X_\infty$, i.e., $X\mapsto X + vt + X_\infty$, such that $\langle X\rangle =0$ in the stationary state. 

As discussed in Sec.~\ref{sec:intro}, the dynamics of the particle can be explored by displacing it from the position of mechanical equilibrium, for example by an amount $X_0$ and by observing the ensuing relaxation. In particular, how does the particle, up to fluctuations, recover its mechanical equilibrium? In the absence of the coupling to the field, the average position of the particle decays exponentially with rate $\omega_0$, independently of $v$. 
In Sec.~\ref{sec:intro} we mentioned that, in viscoelastic media, the relaxation features oscillations for $v\neq 0$, in spite of the overdamped nature of the dynamics.  Remarkably, it turns out that the coupling to the field produces a similar behaviour in the present model \cite{Venturelli_2023}.  This can seen by studying the linearised dynamics of the average position of the particle, given by
\beq
\partial_t \langle X(t)\rangle  = - \langle X(t)\rangle \left[ \omega_0  + \hat \Gamma(s=0)\right] + \int_{-\infty}^t\!\!\rmd t'\, \Gamma(t-t') \langle X(t')\rangle,
\label{eq:eom-lin}
\eeq
where $\Gamma(t)$ is the memory kernel which emerges after the change of reference system and linearisation of the non-linear and non-Markovian force in Eq.~\eqref{eq:eff-force} (see Ref.~\cite{Venturelli_2023} for details),  while $\hat\Gamma(s)$ indicates its Laplace transform. 
Accordingly, for the Laplace transform $\langle \hat X(s)\rangle$ of   $\langle  X(t)\rangle$ one has
\beq
\langle \hat X(s)\rangle  = \frac{X_0}{ s + \omega_0 - \left[ \hat\Gamma(s) - \hat\Gamma(0)\right]} \,,
\label{eq:X-hat}
\eeq
and the possible (damped) oscillations 
of  $\langle  X(t)\rangle$ are related to the emergence of an imaginary part of the zeros of the denominator in this expression. 
This analysis yields the conclusion that, for any value of the dimensionless coupling constant $g\equiv \lambda^2/(\kappa R^d)$ (where $R$ is the radius of the particle, i.e., the length scale associated with the argument of the dimensionless potential $U$), there exist a threshold value of the dimensionless velocity ${\rm Wi} \equiv \tau_R v/R$ above which the relaxation occurs via damped oscillations. In the expression for ${\rm Wi}$, $\tau_R$ stands for the relaxation time of the field $\phi$ at the  wavevector $\pi/R$ associated with the size of the particle (note that $\tau_R$ is finite, even at criticality). In this sense, the dimensionless velocity ${\rm Wi}$ plays the role of the so-called Weissenberg number (see, e.g., Ref.~\cite{Berner_2018}), i.e., the product between the characteristic relaxation rate $\tau_R$ of the medium and its shear rate $v/R$. 
%
%%
%%
%%
%%
%%%%%%%%%%%%%%%%%%%%%%%%%%%%%%%%%%%%%%%%%%
\begin{figure}
    \centering
  \includegraphics[width=0.95\linewidth]{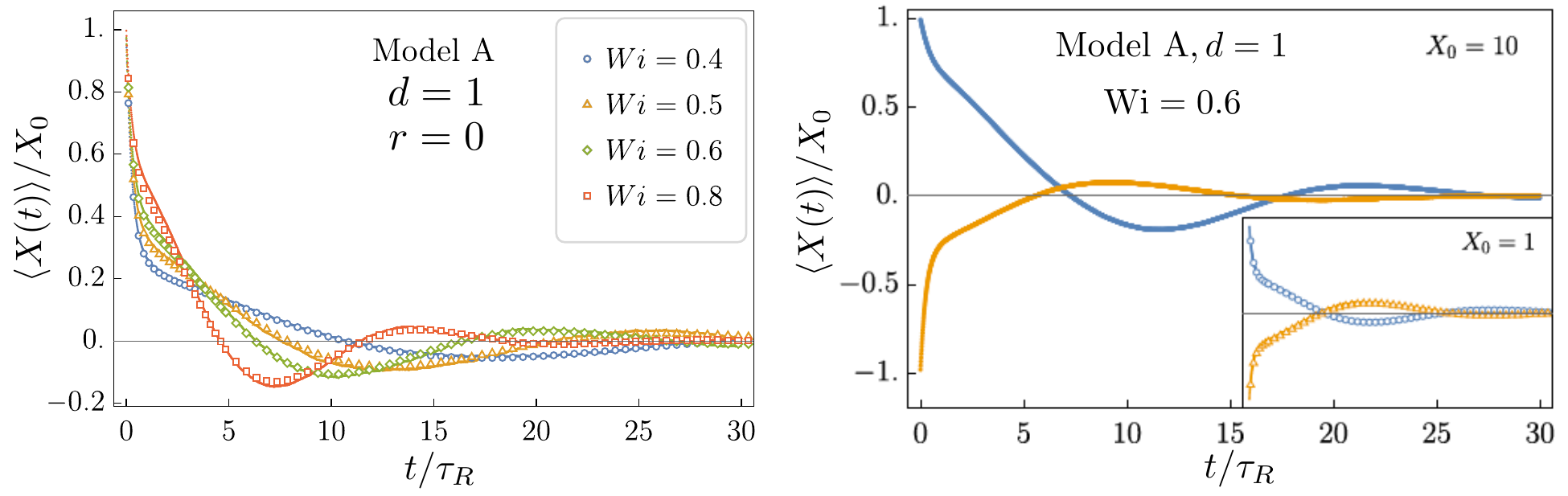}
%   \put(-390,4){(a)}
%  \put(-185,4){(b)}
\caption{Relaxation of the average position $\langle X(t)\rangle$ of the particle in the reference frame co-moving with the trap, after a displacement at $t=0$ of amplitude $X_0$ from the position of mechanical equilibrium in the steady state. The particle interacts with a critical Gaussian field in $d=1$ and model A dynamics.  
Left: Dependence of the relaxation on the dimensionless velocity ${\rm Wi} \propto v$ of the trap. 
The symbols are the analytical predictions within the linearised dynamics, obtained by inverting the Laplace transform in Eq.~\eqref{eq:X-hat}, which are in good agreement with  the result of numerical simulations, indicated by the solid lines. 
The parameters used in the simulations are $\lambda =5$, $R=5$, $\gamma_\infty=0.2$, $\kappa = 0.2$, $D=25$, $X_0=1$, $T=10^{-2}$, and various values of $v$, corresponding to the values of ${\rm Wi}$ indicated in the plot, while $U(x) = \exp(-|x|/R)$.
Right: Dependence of the relaxation of $\langle X(t)\rangle$ on the sign an amplitude $X_0$ of the initial displacement, in a trap with ${\rm Wi = 0.6}$. Solid lines and symbols are obtained as described above. In particular, the blue line and symbols refer to an initial displacement $+X_0$, while the yellow ones to $-X_0$, with $X_0 =10$ in the main plot and $X_0=1$ in the inset.
In both panels, the timescale $\tau_R$ is defined in the text.
(Figure adapted from Ref.~\cite{Venturelli_2023}.)}
\label{fig:dyn-eq-osc}
\end{figure}
%%%%%%%%%%%%%%%%%%%%%%%%%%%%%%%%%%%
%%
%%
%%
%
%
The conclusion above has been drawn on the basis of the linearised dynamics, which is expected to be accurate for $g\ll 1$ or at sufficiently long times. In order to confirm the occurrence of oscillations beyond these cases, we did numerical simulations, the result of which is reported in Fig.~\ref{fig:dyn-eq-osc} for the critical model A in spatial dimension $d=1$.  In particular, the panel on the left shows a comparison between the predictions from the linearised equation (symbols, obtained by inverting the Laplace transform in Eq.~\eqref{eq:X-hat}) and the results of numerical simulations of the evolution equation for the particle and the field (solid lines). 
The numerical simulations confirm the theoretical predictions both quantitatively and qualitatively for the various values of the speed $v$.
These oscillations emerge as a consequence of the memory in the dynamics of the particle which is induced  by the combination of the presence of the field and of the drive. In passing, we note that the memory kernel in Eq.~\eqref{eq:eom-lin} features correlations at short times and anti-correlations at longer times, which resembles those in Eq.~\eqref{eq:Gamma-fit}, although the physical mechanism for the emergence of oscillations differs from the one in viscoleastic media (see Ref.~\cite{Venturelli_2023} for details).
As it has been emphasised in Sec.~\ref{sec:model}, the effective equation of motion for the particle is non-linear: this affects, for example, the statistic of the stationary displacement discussed in Sec.~\ref{sec:eq-dyn} and it gives rise to a clear pre-asymptotic regime with algebraic decay in the relaxation of the particle in the trap for $v=0$ \cite{Venturelli_2022}. An alternative way of highlighting the role of non-linearities in the effective dynamics of the driven particle is to consider its relaxation starting from a position with a displacement $\pm X_0$. In fact, if the effective equation is linear, the evolution of $\langle X(t)\rangle$ starting from $X_0$ is the opposite of the one starting from $-X_0$. For the driven particle this is actually \emph{not} the case, as shown in the right panel of Fig.~\ref{fig:dyn-eq-osc}, for model A dynamics in $d=1$ and at criticality and with a certain drag velocity ${\rm Wi}$. In fact, the blue and yellow curves, which differ only for the sign of the initial displacement $\pm X_0$, cannot be obtained one from the other by a reflection $X \leftrightarrow - X$. However, upon decreasing the value of $X_0$, one expects the linear approximation to become increasingly more accurate and thus this symmetry to be restored: this is shown by the curves in the inset, which refer to an initial displacement $X_0$ ten times smaller than the one considered in the main plot and which are indeed related to each other by a space reflection.
%

%%
%%
%%
%%
%%%%%%%%%%%%%%%%%%%%%%%%%%%%%%%%%%%%%%%%%%
\begin{figure}
    \centering
  \includegraphics[width=0.57\linewidth]{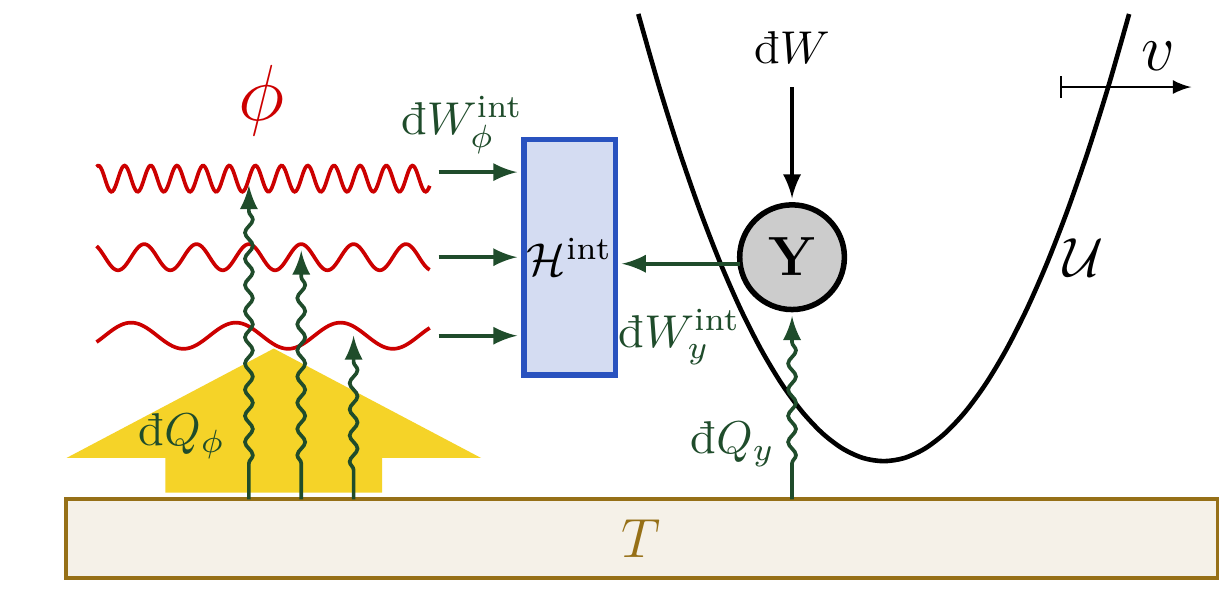}
%   \put(-390,4){(a)}
%  \put(-185,4){(b)}
\caption{
Sketch of the relevant degrees of freedom which intervene in the stochastic thermodynamics of the system consisting of the field $\phi$ in interaction with a particle $\mathbf{Y}$ subject to a time-dependent potential ${\cal U}$, with the particle and the field in contact with an equilibrium thermal bath at temperature $T$. The field-particle interaction ${\cal H}^{\rm int}$ (see the text) can store energy. In particular, the particle and the field exchange heats $\dbar Q_y$ and $\dbar Q_\phi$ with the thermal bath, while the external force exerts a work $\dbar W$ on the particle. The particle and the field exchange energies $\dbar W_y^{\rm int}$ and $\dbar W_\phi^{\rm int}$, respectively, via their interaction, which may also change ${\cal H}^{\rm int}$. The modes of the field are indicated by the red waves and, for clarity, the exchanges of work among them
are not indicated in the sketch. Heat supplied to the particle and field by the thermal bath is conventionally assumed to be positive.
(Figure adapted from Ref.~\cite{venturelli2023stochastic}.)}
\label{fig:st-dof}
\end{figure}
%%%%%%%%%%%%%%%%%%%%%%%%%%%%%%%%%%%
%%
%%
%%
%%
%%

\subsection{Stochastic thermodynamics}
\label{sec:s-td}

Another interesting aspect of the spatially extended system constituted by the particle in interaction with the field, under the effect of the external force, and in contact with the thermal bath is its stochastic thermodynamics. In particular, its investigation requires an extension \cite{venturelli2023stochastic} of the standard approach (see, e.g., Refs.~\cite{PP_2021,S_2025}) to the case in which the medium is characterised by spatio-temporal correlations. In practice, the various modes of the field, represented as red waves in  Fig.~\ref{fig:st-dof}, are additional degrees of freedom in interaction among themselves (exchanging work) but also with both the bath (lower rectangle), with which they exchange heat, and the particle, denoted by $\mathbf{Y}$ and represented by a grey circle. This interaction with the particle occurs via the interaction Hamiltonian ${\mathcal H}^{\rm int}$ (blue rectangle) --- corresponding to the last term in Eq.~\eqref{eq:model} --- which can store energy. The particle, in turn, is also subject to a time-dependent external potential ${\cal U}$ (the moving trap of the previous Sec.~\ref{sec:driving}) and exchanges heat with the bath. Both the particle and the modes of the field do a space-dependent work $\dbar W^{\rm int}_y(x)$ and $\dbar W^{\rm int}_\phi(x)$, respectively, which change locally ${\mathcal H}^{\rm int}$. Similarly, the heat $\dbar Q_\phi(x)$ and $\dbar Q_y(x)$ extracted from the bath by the field $\phi$ and the particle, respectively, depend on space. These heat and work fields can be eventually expressed in terms of  the quantities in Eqs.~\eqref{eq:dynamics-p-f} and \eqref{eq:noises}, as discussed in Ref.~\cite{venturelli2023stochastic}.

\begin{figure}
    \centering
  \includegraphics[width=0.95\linewidth]{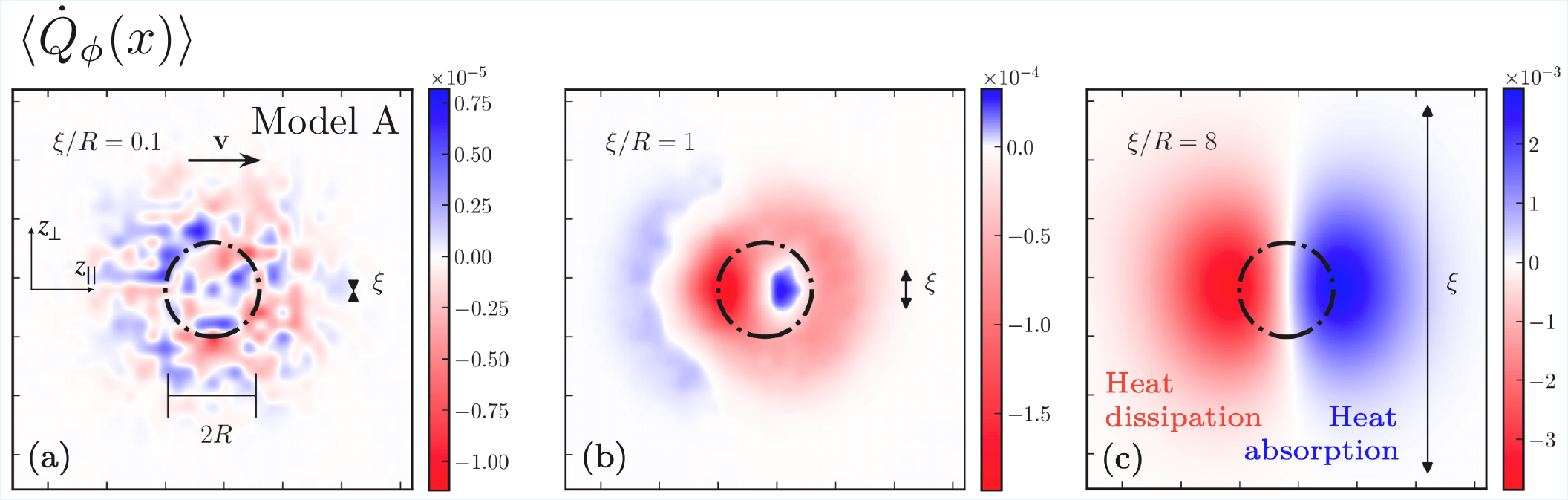}
%   \put(-390,4){(a)}
%  \put(-185,4){(b)}
\caption{Spatial distribution of the heat absorption rate $\langle \dot{Q}_\phi(x)\rangle$ of the field from the thermal bath, in a frame co-moving with the particle driven at velocity $v$ along the horizontal direction, in spatial dimensions $d=2$. The particle is driven horizontally to the right with velocity $\mathbf{v}$, while the Gaussian field evolves with model A dynamics. The particle is assumed to have $U(x) = \exp(-x^2/(2R))$ with radius $R=4$, indicated by the dash-dotted lines in the various panels.  The heat map is the result of the numerical simulation of the dynamics of the system (see Ref.~\cite{venturelli2023stochastic} for details), with an increasing value of the ratio $\xi/R$ from 0.1 on the left to 8 on the right, where $\xi$ is the correlation length, indicated by the arrows in the panels. Upon increasing $\xi$, a clear ``dipolar'' structure emerges in  $\langle \dot{Q}_\phi(x)\rangle$, with heat absorption in front of the particle and heat dissipation in its back.  (Figure adapted from Ref.~\cite{venturelli2023stochastic}.) }
\label{fig:st-heat}
\end{figure}

The consequences of having correlated thermal fluctuation within the medium can be appreciated, for example, by focussing on the heat $\dbar Q_\phi(x)$ that the field $\phi$ extracts in a small time interval ${\rm d}t$ from the thermal bath, i.e., on the average rate $\langle \dot{Q}_\phi\rangle = \langle\dbar Q_\phi(x) \rangle/{\rm d}t $ indicated by the yellow arrow in Fig.~\ref{fig:st-dof}.
For a particle in spatial dimensions $d=2$, which is driven along the horizontal direction with a certain velocity $v$, the heat map of $\langle \dot{Q}_\phi\rangle$, as obtained in the co-moving frame from numerical simulations, is reported in Fig.~\ref{fig:st-heat}. The field  $\phi$ is assumed to evolve with model A dynamics and its correlation length $\xi$, indicated by an arrow in the panels, grows from left to right. The boundary of the particle is indicated by the dash-dotted line. Interestingly enough, one observes that, as $\xi$ grows and exceeds $R$, a well-defined spatial organisation of  $\langle \dot{Q}_\phi\rangle$ emerges from the initial structureless and tiny spatial fluctuations (left panel), until a sort of ``dipole'' forms (right) in which $\phi$, on average, extracts heat from the bath in front of the moving particle, while it dissipates it in the bath behind the particle.  The emergence of such a structure is a sole consequence of correlations within the medium, which allow the particle to have effects on the medium also rather far from its center. 
%%
%%
%%%%%%%%%%%%%%%%%%%%%%%%%%%%%%%%%%%%%%%%%%
%%%%%%%%%%%%%%%%%%%%%%%%%%%%%%%%%%%
%%
%%
%%

\section{Making the particle active} 
\label{sec:active}

\begin{figure}
    \centering
  \includegraphics[width=0.65\linewidth]{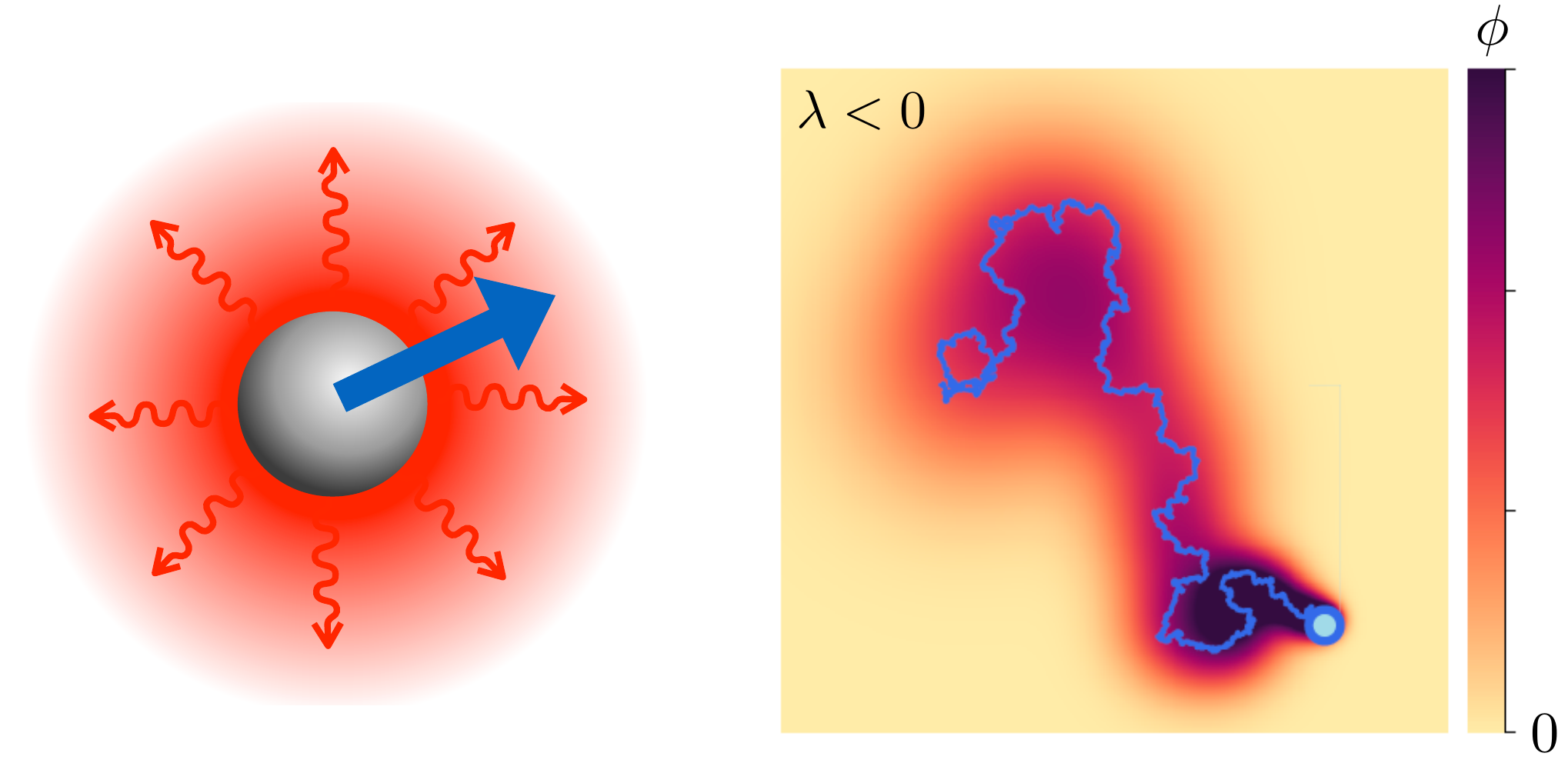}
%   \put(-390,4){(a)}
%  \put(-185,4){(b)}
\caption{Left: Cartoon of a self-chemotactic particle which releases in the environment around it a chemical (red) with diffusing density $\phi$ (wiggly arrows). The particle is then subject to a force (blue arrow) aligned with the gradient of $\phi$ which it experiences. Right: Sample trajectory (blue) of a self-chemotatic particle in spatial dimension $d=2$, with repulsive chemotaxis $\lambda <0$ which makes it run away from regions in space where $\phi$ is larger. The heat map indicates the local value of $\phi$ at the time at which the particle is at the final point of its trail, marked by a blue dot.}
\label{fig:rtm}
\end{figure}

The model presented in Sec.~\ref{sec:model} lends itself to an interesting generalisation out of equilibrium, which might even render the particle active \cite{Romano_2026}.
In the previous Sections, the interaction between the particle and the field $\phi$ was reciprocal --- as it derived from an interaction in the common Hamiltonian ${\cal H}$ --- and the noises satisfied the fluctuation-dissipation relations.
A possible way of violating the latter conditions is to assume, for example, that no fluctuations affect the evolution of the field $\phi$ (i.e., that the noise $\eta$ in Eq.~\eqref{sec:model} vanishes). 
Moreover, one can also break the reciprocity of the interaction between the field $\phi$ and the particle, by fixing, in their equations of motion \eqref{eq:dynamics-p-f}, the corresponding interaction strengths (which, in equilibrium, are both given by $\lambda$) to different values in the two equations.
As a result, these interactions can no longer be derived by a common interaction Hamiltonian ${\cal H}^{\rm int}$. When this is done starting from the model in Eq.~\eqref{sec:model}, with the field evolving at criticality with model A dynamics instead of the model B considered there, and with a point-like interaction $U(x) = \delta^d(x)$, one arrives at the following non-equilibrium,
non-reciprocal model upon setting $\lambda=1$ solely in the equation of motion for the field: 
\beq
\begin{cases}
\partial_t \phi(x,t) =  D \nabla^2\phi(x,t) + \delta^d(x-X),\\
\underbrace{m \ddot{X}}_{= 0} = -\gamma_\infty \dot{X}  + \lambda \grad \phi(X,t) + \zeta(t).
\end{cases}
\label{eq:selfchemo}
\eeq
The correlations of the zero-mean Gaussian noise $\zeta$ are given by the second line of Eq.~\eqref{eq:noises}.
Actually, these equations of motion can be easily interpreted as describing the phenomenon of \emph{self chemotaxis} in which the particle actually moves in response to the chemical gradient that it generates itself, as shown by the cartoon in the left panel of Fig.~\ref{fig:rtm}. 
In fact, the first equation  in Eq.~\eqref{eq:selfchemo} describes the diffusion of the ``chemical'' with spatial density $\phi$, which has the particle at position $X$ as its $\delta$-source. In practice, the particle releases a chemical which diffuses in the environment, as sketched in the figure. The second equation, instead, indicates that the particle is subject not only to the thermal noise $\zeta$ but also to the force $\lambda \nabla \phi(X,t)$ (indicated by an arrow in the left panel of Fig.~\ref{fig:rtm}) which drives it towards regions in space with larger or smaller values of $\phi$ depending on $\lambda$ being positive or negative. As the particle is also the source of field, this means that it tends to escape from itself for $\lambda<0$, giving rise to the repulsive chemotaxis, as opposed to the attractive one for $\lambda>0$. 
Note that, in general, one has to regularise the equations above because the gradient of the field at the same position 
in which it is generated is generically not well defined. 
%%
%%
%%
%%
%%%%%%%%%%%%%%%%%%%%%%%%%%%%%%%%%%%%%%%%%%
%%%%%%%%%%%%%%%%%%%%%%%%%%%%%%%%%%%
%%
%%
%%
For repulsive chemotaxis $\lambda <0$, the particle tends to escape from its own trail: in this respect, it is not surprising that, in the absence of the noise, the deterministic part of Eq.~\eqref{eq:selfchemo}, admits a solution with a spontaneous velocity $v\propto -\lambda$ 
that can be predicted analytically~\cite{michelin_spontaneous_2013, chamolly_stochastic_2019, morozov_self-propulsion_2019,picella_confined_2022,Romano_2026}: this renders the particle \emph{active} and self-propelled. 
In the presence of the noise, this velocity of self propulsion is perturbed, making the dynamics resemble a run-and-tuble process. For illustration, the right panel of Fig.~\ref{fig:rtm} shows the particle trajectory (blue) in two spatial dimensions, while the heat map indicates the value of the field $\phi$ as a function of space, at the time at which the particle is in the final position of its trail, indicated by the blue dot. 

In order to determine the consequences of this self chemotaxis on the dynamics of the particle, it is convenient to focus on the 
mean square displacement ${\rm MSD} = \langle [X(t)-X(0)]^2\rangle$ when the particle and the field are considered in $d$ spatial dimensions.
If the chemical emitted by the colloid decays in time (which corresponds to adding a term $-\kappa \phi$ on the r.h.s.~of the evolution equation for $\phi$ in Eq.~\eqref{eq:selfchemo}), the particle quickly loses memory of its past motion: velocity correlations decay rapidly and the long-time dynamics is diffusive, with a renormalized diffusion coefficient~\cite{grima_phase_2006,grima_strong-coupling_2005}. In contrast, for the chemically persistent species produced by most chemotactic particles, the self-generated chemical landscape retains a record of the particle's trajectory, leading to strong nonlinear memory effects in the dynamics. 
While previous studies were primarily concerned with determining 
effective diffusion coefficients in this case
(see, e.g., Refs.~\cite{sengupta_dynamics_2009,daftari_self-avoidant_2022}),
here we focus on the long-time behaviour of the MSD.
It actually turns out \cite{Romano_2026} that  a sort of upper critical
dimensionality $d_c$ emerges, with $d_c=2$, such that normal diffusion with ${\rm MSD} \propto t$ as $t\to\infty$ is recovered at long times for  $d>d_c$, independently of whether the chemotaxis is repulsive or attractive (i.e., independently of the sign of $\lambda$).
 For $d=d_c$, the MSD develops, as expected, logarithmic corrections as a function of time. For $d<d_c$ and, in particular in $d=1$, the dynamics becomes  superdiffusive with ${\rm MSD} \propto t^{4/3}$  for $\lambda<0$, while the dependence on time is logarithmic  ${\rm MSD} \propto \log t $ for $\lambda>0$ because the trail of the particle hinders its motion by attracting it.  
In the repulsive case $\lambda <0 $, the superdiffusive behavior is due to the fact that, as anticipated, the motion of the particle resembles a run-and-tumble motion, where the run occurs approximately with the self-propulsion velocity $v$, while the noise determines a tumbling process with a strong memory given by the field that the particle generates while moving. The predictions described above can be derived analytically, in part with suitable scaling arguments and, for the repulsive case in $d=1$, by determining from Eq.~\eqref{eq:selfchemo} an effective run-and-tumble process with correlated tumbles which, in turn, can be solved analytically. These predictions were also confirmed by numerical simulations \cite{Romano_2026}.  
We also mention that the repulsive chemotaxis in $d=1$ turns out to be related to the 
so-called true self-avoiding random walk (TSAW), which was studied long ago in Refs.~\cite{amit_asymptotic_1983,pietronero_critical_1983,obukhov_renormalisation_1983}. 

\section{Conclusions and perspectives} 
\label{sec:outlook}

The simple model introduced in Sec.~\ref{sec:model} lends itself to a number of interesting and relevant extensions, that may also bring it closer to realistic experimental systems. Among them, we mention: 
\begin{itemize}
\item[(1)] Explore additional and observable consequences of the memory emerging in the correlated medium. For example, Ref.~\cite{Pruszczyk_2025} demonstrated that the driven particle discussed in Sec.~\ref{sec:driving} may feature a memory-induced recoil after switching off the drive, analogously to what was observed  in viscoelastic media \cite{Ginot_2022_rec}.
Similarly, motivated by recent investigations in this kind of media \cite{Ginot_2022,Cao_2023,Gomez-Solano_2014,Jain_2021_2}, it would be interesting to study the effects of correlations on, e.g., microrheology (partly addressed in Ref.~\cite{venturelli2023stochastic}) or the influence on barrier crossing etc. 
\item[(2)] Account for the self-interaction of the field $\phi$, e.g., $\propto \phi^4$ is certainly necessary in order to be able to explore the connection with actual critical systems, where it is well-known that the fluctuations of the order parameter field $\phi$ are non-Gaussian. In this direction, it would be important to consider  models of field dynamics which are more realistic for actual fluids such as, e.g., model H \cite{Halperin_1977}. 
Similarly, the extension to the case in which the coupling between the particle and the field is also non-linear in the field might be or relevance for certain applications. 
\item[(3)] Considering the simultaneous presence of more than one particle within the medium opens up the possibility to explore the effects of correlation  on the dynamics of the particles and on the possible structure that they may form. This is particularly relevant for understanding, e.g., the dynamics of fluctuation-induced forces beyond mean-field theory \cite{PhysRevLett.111.055701} or the dynamics of polymers \cite{Muzzeddu_2025} in such media. In this direction, 
Ref.~\cite{Pruszczyk_2025b}   has recently studied the model of Sec.~\ref{sec:model} in which the field $\phi$ is constrained to vanish at a flat boundary, reminding the classical setting in which Casimir-like forces emerge \cite{Gambassi_2009}. The particle then interacts with $\phi$ as in Eq.~\eqref{eq:model} and, because of the constraint on $\phi$, it turns out to be subject to a repulsive force from the boundary, with a range set by the correlation length $\xi$. 

\item[(4)] Considering the case in which the particle is active in a passive medium at equilibrium or the complementary case in which the particle is passive but the underlying medium is active is also of practical and conceptual interest.
\end{itemize}

\section*{Acknowledgments}
Several of the results mentioned in this article were obtained in collaboration with Urna Basu,  Vincent D\'emery,  Francesco Ferraro,  Sarah Loos,  Pierluigi Muzzeddu,  Marcin Pruszczyk, \'Edgar Rold\'an, Jacopo Romano, Davide Venturelli, and Benjamin Walter to whom I am really grateful.  I also would like to thank S.~Dietrich, Clemens Bechinger, Matthias Kr\"uger and Sergio Ciliberto for invaluable discussions and Jacopo Romano for a careful reading of the manuscript.

\section*{References}
\providecommand{\newblock}{}

\end{document}